\documentstyle[aps]{revtex}
\input{tcilatex}

\begin{document}
\title{Form factors for $B\longrightarrow \pi l\nu $ decay in a model constrained
by chiral symmetry and quark model}
\author{Riazuddin \thanks{%
E-mail address: riazuddins@yahoo.com}}
\address{Department of Physics, King Fahd University of Petroleum and Minerals,\\
Dhahran 31261, Saudi Arabia \\
and National Center for Physics, Quaid-e-Azam University, Islamabad 45320,\\
Pakistan.}
\author{T.A. Al-Aithan \thanks{%
E-mail address: alaithan@kfupm.edu.sa}}
\address{Department of Physics, King Fahd University of Petroleum and Minerals,\\
Dhahran 31261, Saudi Arabia.}
\author{Amjad Hussain Shah Gilani \thanks{%
E-mail address: ahgilani@yahoo.com}}
\address{Department of Physics and National Center for Physics, Quaid-e-Azam\\
University, Islamabad 45320, Pakistan.\\
Preprint: NCP-QAU/0007-15}
\maketitle

\begin{abstract}
The form factors for the $B\longrightarrow \pi $ transition are evaluated in
the entire momentum transfer range by using the constraints obtained in the
framework combining the heavy quark expansion and chiral symmetry for light
quarks and the quark model. In particular, we calculate the valence quark
contributions and show that it together with the equal time commutator
contribution simulate a $B-$meson pole $q^2-$dependence of form factors in
addition to the usual vector meson $B^{*}-$pole diagram for $%
B\longrightarrow \pi l\nu $ in the above framework. We discuss the
predictions in our model, which provide an estimate of $\left| V_{ub}\right|
^2$. PACS number(s): 13.20.-V, 12.39.Ki
\end{abstract}

\section{Introduction}

There has been great interest in the study of semileptonic decays of heavy
mesons as they provide a testing ground for heavy-quark effective theory
(HQET). The symmetry underlying this theory allows to derive model
independent predictions for the form factors near the zero recoil point. On
the other hand for heavy -to-light semileptonic transitions, there exists no
symmetry principle to guide us. However, here the heavy quark expansion can
be combined with chiral symmetry and PCAC for final pseudoscalar mesons \cite
{r1,r2,r3,r4,r5}. We focus on $B\longrightarrow \pi l\nu $ which is
important for the evaluation of the CKM matrix element $V_{ub}$.

There have been many calculations \cite
{r1,r2,r3,r4,r5,r6,r7,r8,r9,r10,r11,r12,r13,r14,r15,r16,r17,r18,r19,r20,r21,r22,r23,r24,r25,r26}
for $B\longrightarrow \pi l\nu $ form factors, using different approches, in
the past. In this paper we follow the approach used in ref. 4. In $%
B\longrightarrow \pi l\nu $ decay , the vector meson pole $B^{*}$, with mass
degenerate with $B$ in the heavy quark symmetry limit, dominates the
transition amplitude at the zero recoil point. We calculate the valence
quark contribution, and find that it together with the equal-time
contribution is as important as the $B^{*}-$meson pole dominance of the form
factors since the former two simulate a $B-$meson pole like $q^2$ dependence
of the form factors. We perform this calculation in a framework compatable
with chiral symmetry and eliminate the $B$-meson bound state function in
favor of $B$-meson decay constant $f_B$ which can likewise be calculated in
the valance quark approximation. This procedure gives us form factors for $%
q^2$ in the range determined by $E_\pi <\Lambda \ \left( \sim 1\text{ GeV}%
\right) ,$ where $\Lambda $ is some interaction scale, below which chiral
symmetry should be valid. This constraint is then built into an
extrapolation function for $f_{+}\left( q^2\right) $ which determines $%
f_{+}\left( q^2\right) $ in the entire $q^2-$range. Finally, we compare our
results with some of the earlier calculations and also obtain an estimate of 
$\left| V_{ub}\right| $ by using CLEO data.

\section{Current Algebra Constraints}

The relevant hadronic matrix elements for the $B\rightarrow \pi l\nu $ decay
is defined as 
\begin{eqnarray}
T_\mu &=&\left\langle \pi ^{+}(k)|(V_\mu )|B^0(P)\right\rangle  \nonumber \\
&=&f_{+}(t)(P+k)_\mu +f_{-}(t)q_\mu  \label{1}
\end{eqnarray}
with $V_\mu $ the weak vector current $V_\mu =\bar{u}\gamma _\mu b$ and $%
q=p-k,t=q^2$. Following Ref. 4, the current algebra constraint at $k^2=0$
(but not $k=0)$ is 
\begin{eqnarray}
\widetilde{T}_\mu =-\left( \frac{ik^\lambda }{f_\pi }M_{\lambda \mu
}^{B^{*}}+T_\mu ^{B^{*}}\right) +\frac{f_B}{f_\pi }p_\mu -i\frac{k^\lambda }{%
f_\pi }\widetilde{M}_{\lambda \mu }  \label{2}
\end{eqnarray}
where the tilde denotes that $B^{*}$ pole terms have been seperated out from 
$T_{\mu \text{ }}$and from $M_{\lambda \mu }$ respectively, and $f_\pi
=0.132GeV.$ The third term on the right hand side of Eq.(\ref{2}) comes from
the equal time commutator. $M_{\lambda \mu }$ is defined as 
\begin{eqnarray}
M_{\lambda \mu }=i\int d^4x\,\exp (ik\cdot x)\ \left\langle 0\left| T\left(
A_\lambda ^{1-i}(x),\,V_\mu (0)\right) \right| B^0(P)\right\rangle  \label{3}
\end{eqnarray}
where $A_\lambda ^{1-i2}$ is the axial vector current $\bar{d}\gamma
_\lambda \gamma _5u$. The term within the parenthesis in Eq. (\ref{2}) is
easily evaluated to be $-g_{BB\pi }f_{B^{*}}/m_B^2$ and one obtains 
\begin{eqnarray}
T_\mu &=&T_\mu ^{B^{*}}+\widetilde{T}_\mu  \nonumber \\
&=&\frac{g_{B^{*}B\pi }f_{B^{*}}}{M_{B^{*}}^2-t}\left[ (P+k)_\mu -\frac{%
M_B^2-m_\pi ^2}{M_{B^{*}}^2}q_\mu \right] -\frac{g_{B^{*}B\pi }f_{B^{*}}}{%
M_{B^{*}}^2}q_\mu +\frac{f_B}{f_\pi }\frac 12\left[ (P+k)_\mu +q_\mu \right]
-\frac{ik^\lambda }{f_\pi }\widetilde{M}_{\lambda \mu }  \label{4}
\end{eqnarray}
where one can identify $\widetilde{M}_{\lambda \mu }$ as a source of
corrections to the leading $B^{*}$ pole contribution. Next we identify such
corrections, as valence quark conribution to $\widetilde{M}_{\lambda \mu }$
which we evaluate in the next section.

\section{Valence Quark Contribution}

The valence quark contribtion is shown in Figure 1. Its evaluation gives 
\begin{eqnarray}
\widetilde{M}_{\lambda \mu }=i\int \frac{d^3K}{(2\pi )^3}A_{\lambda \mu }
\label{5}
\end{eqnarray}
where $A_{\lambda \mu }$ is the matrix element of Figure 1: 
\begin{eqnarray}
A_{\lambda \mu } &=&-i\left( \sqrt{2M_B}\frac 1{\sqrt{2}}\sqrt{3}\phi _B(%
{\bf K)}\bar{u}^i(p_b)(\gamma _5)_i^jv(p_d)_j\right) \sqrt{\frac{m_d}{p_{d_0}%
}}\bar{v}^r(p_d)(\gamma _\lambda \gamma _5)_r^s\frac{(m_u+{p\hspace{-0.2cm}/}%
_u)_s^m}{p_u^2-m_u^2}(\gamma _\mu )_m^nu_n(p_b)\sqrt{\frac{m_b}{p_{b_0}}} 
\nonumber \\
&&  \label{6}
\end{eqnarray}
In equation (\ref{6}), the term within the parenthesis is the bound state
wave function of B-meson, $\sqrt{3}$ being the color factor. We define the
variables ${\bf K}={\bf p}_b-{\bf p}_d,\,{\bf P}={\bf p}_b+{\bf p}_d$, so
that ${\bf K}$ is the relative momentum and ${\bf P}$ is the centre of mass
mometum of the $b\bar{d}$ system. Eq.(\ref{6}) gives 
\begin{eqnarray}
A_{\lambda \mu } &=&-i\sqrt{2m_B}\frac 1{\sqrt{2}}\sqrt{3}\sqrt{\frac{m_dm_b%
}{p_{do}p_{bo}}}\frac 1{4m_bm_d}\phi _B({\bf K})\frac 1{p_u^2-m_u^2} 
\nonumber \\
&&\times \left\{ -Tr(\gamma _\lambda \gamma _5)(m_u+{p\hspace{-0.2cm}/}%
_u)\gamma _\mu (m_b+{p\hspace{-0.2cm}/}_b)\gamma _5({p\hspace{-0.2cm}/}%
_d-m_d)\right\}  \nonumber \\
&&  \label{7}
\end{eqnarray}

The above $-Tr$ is evaluated to be $(m_u=m_d=m_q)$ 
\begin{eqnarray}
4\left\{ g_{\lambda \mu }(m_q^2m_b-m_qp_u\cdot p_b+m_qp_b\cdot
p_d-m_bp_u\cdot p_d)-m_b(p_{u\mu }p_{d\lambda }+p_{u\lambda }p_{d\lambda
})-m_qp_{b\lambda }(p_u-p_d)_\mu -m_qp_{b\mu }(p_{u+}p_d)_\lambda \right\}
\label{8}
\end{eqnarray}
Working in the rest frame of $B$-meson $\left( {\bf P}=0\right) $ where 
\begin{eqnarray*}
\left( \frac{{\bf K}^2}4+m_b^2\right) ^{1/2} &=&\frac{M_B^2+\left(
m_b^2-m_q^2\right) }{2M_B}, \\
K_0 &=&\frac{m_b^2-m_q^2}{2M_B}.
\end{eqnarray*}
and 
\begin{eqnarray}
p_u^2-m_u^2=\frac{M_B^2-m_b^2+m_q^2}2\left( 1-\frac{q^2}{M_B^2}\right) -{\bf %
q}\cdot {\bf K}  \label{9}
\end{eqnarray}
\begin{eqnarray}
-ik^\lambda A_{\lambda \mu }=4C({\bf K})\frac{-1}{L^{\prime }+{\bf q}\cdot 
{\bf k}}\left\{ (a+b{\bf q}\cdot {\bf K})k_\mu +(a^{\prime }-{b}{\bf q}\cdot 
{\bf K})(K-q)_\mu \right\}  \label{10}
\end{eqnarray}

Here 
\begin{eqnarray}
C({\bf K}) &=&\sqrt{2M_B}\frac 1{\sqrt{2}}\sqrt{3}\sqrt{\frac{m_dm_b}{%
p_{d_0}p_{b_0}}}\frac 1{4m_bm_d}\phi _B(K),  \label{11} \\
L^{\prime } &=&-\frac{M_B^2-m_b^2+m_q^2}2\left( 1-\frac{q^2}{M_B^2}\right)
\label{12} \\
a &=&\frac{M_B^2-(m_b-m_q)^2}2\left\{ -\frac 12(m_b+m_q)\left( 1-\frac{q^2}{%
M_B^2}\right) +2m_q\right\}  \label{13} \\
b &=&\frac 12(m_b-m_q)  \label{14} \\
a^{\prime } &=&\frac 14\left( M_B^2-(m_b-m_q)^2\right) (m_b+m_q)\left( 1-%
\frac{q^2}{M_B^2}\right)  \label{15}
\end{eqnarray}
When Eq.(\ref{10}) is put in (\ref{5}) and the angular integration is
carried out, one gets for example, 
\begin{eqnarray}
2\pi \int_{-1}^1\frac{a+b{\bf q}\cdot {\bf K}}{L^{\prime }+{\bf q}\cdot {\bf %
K}}=2\pi \left\{ \frac{a-bL^{\prime }}{|{\bf q}||{\bf K}|}\ln \frac{%
L^{\prime }+|{\bf q}||{\bf K}|}{L^{\prime }-|{\bf q}||{\bf K}|}+2b\right\}
\label{16}
\end{eqnarray}
Then, noting that if $\phi _B({\bf K})$ is of Gaussian type, $|{\bf K}%
|\simeq 0$ dominates in the $K-$integration, so that we can expand the
logarithm in Eq.(\ref{16}) and thus this equation reduces to 
\begin{eqnarray}
2\pi \left\{ \frac{a-bL^{\prime }}{|{\bf q}||{\bf K}|}\,\frac{2|{\bf q}||%
{\bf K}|}{L^{\prime }}+2b\right\} =4\pi \frac a{L^{\prime }}  \label{17}
\end{eqnarray}
Further $4\pi \int K^2dK\,\phi _B(K)$ becomes $\int d^3K\,\phi _B(K),$ which
is the Fourier transform of the wave function at the origin which we write
as $\phi _B(0).$ As far as the intergation involving $K_\mu $ is concerned,
it is easy to see that the angular integration involving ${\bf K}$ gives
zero while that over $K_0$ gives $4\pi K_0=4\pi \frac{m_b^2-m_q^2}{M_B^2}M_B$
so that in the rest frame of B-meson the angular integration involving $%
K_\mu $ gives $4\pi \frac{m_b^2-m_q^2}{M_B^2}P_\mu $. Thus finally we obtian 
\begin{eqnarray}
-ik^\lambda \widetilde{M}_{\lambda \mu }=-4C(0)\left[ \frac a{L^{\prime }}%
k_\mu +\frac{a^{\prime }}{L^{\prime }}\left( \frac{m_b^2-m_q^2}{M_B^2}P_\mu
-q_\mu \right) \right]  \label{18}
\end{eqnarray}
where 
\begin{eqnarray}
C(0)=\sqrt{2M_B}\frac 1{\sqrt{2}}\sqrt{3}\frac 1{4m_bm_q}\phi _B(0)
\label{19}
\end{eqnarray}
Thus, writing $k_\mu =\left( \frac{P+k}2\right) _\mu -\frac 12q_\mu ,\,P_\mu
=\left( \frac{P+k}2\right) _\mu +\frac 12q_\mu ,$ we finally obtain the
valence quark contribution to the form factors $f_{\pm }$ as 
\begin{eqnarray}
f_{+}^{\text{valence}}(q^2) &=&-\frac{4C(0)}{2f_\pi L^{\prime }}\left\{
a+\left( \frac{m_b^2-m_q^2}{M_B^2}\right) a^{\prime }\right\}  \label{20} \\
f_{-}^{\text{valence}}(q^2) &=&\frac{4C(0)}{2f_\pi L^{\prime }}\left\{
a+\left( \frac{2M_B^2-m_b^2+m_q^2}{M_B^2}\right) a^{\prime }\right\}
\label{21}
\end{eqnarray}

To eliminate $4C(0)$, we consider the matrix elements 
\begin{eqnarray}
\left\langle 0\left| A_\lambda \right| B(p)\right\rangle =if_Bp_\lambda
\label{22}
\end{eqnarray}
which, when evaluated in the same valence quark approximation employed for
the calculation of $-ik_{\lambda \mu }^\lambda \widetilde{M}_{\lambda \mu }$%
, give 
\begin{eqnarray}
f_B=\frac{4C(0)}{2M_B^2}(m_b+m_q)\left[ M_B^2-(m_b-m_q)^2\right]  \label{23}
\end{eqnarray}
Then, using Eqs.(\ref{11}--\ref{13}, \ref{15}) and (\ref{23}), we obtain
from Eq.(\ref{20}, \ref{21}) 
\begin{eqnarray}
f_{+}^{\text{valence}}(q^2) &=&\frac{f_B}{2f_\pi }\left\{ -1+\frac{4m_qM_B^2%
}{(m_b+m_q)\left( M_B^2-(m_b-m_q)^2\right) }\frac 1{1-q^2/M_B^2}\right\}
\label{24} \\
f_{-}^{\text{valence}}(q^2) &=&-\frac{f_B}{2f_\pi }\left\{ 1+\frac{4m_qM_B^2%
}{(m_b+m_q)\left( M_B^2-(m_b-m_q)^2\right) }\frac 1{1-q^2/M_B^2}\right\}
\label{25}
\end{eqnarray}

\section{Combined Contribution To $f_{+}(q^2)$ in the chiral symmetry limit}

Using Eq. (\ref{4}) $\left[ t=q^2\right] ,$ we obtain $\left[
f_{B^{*}}=M_Bf_B\right] $ \cite{r4} 
\begin{eqnarray}
f_{+}(t) &=&\frac{f_B}{2f_\pi }+f_{+}^{\text{valence}}(t)+\frac{g_{B^{*}B\pi
}f_BM_B}{M_{B^{*}}^2-t}  \label{26} \\
f_{-}(t) &=&\frac{f_B}{2f_\pi }+f_{-}^{\text{valence}}(t)+\frac{g_{B^{*}B\pi
}f_BM_B}{M_{B^{*}}^2}-\frac{M_B^2-m_\pi ^2}{M_{B^{*}}^2}\frac{g_{B^{*}B\pi
}f_BM_B}{M_{B^{*}}^2-t}  \label{27}
\end{eqnarray}
The coupling constant $g_{B^{*}B\pi \text{ }}$has been parametrized as 
\begin{eqnarray}
g_{B^{*}B\pi \text{ }}=\frac{\lambda M_B}{f_\pi }  \label{32}
\end{eqnarray}
where $\lambda $ lies in the range $0.3\leq \lambda \leq 0.7$.

We combine the equal time contribution $\frac{f_B}{2f_\pi }$ with $f_{\pm }^{%
\text{valence}}\left( q^2\right) $ and call it continum contribution: 
\begin{eqnarray}
f_{\pm }^{\text{cont.}}\left( q^2\right) =\pm \lambda _c\frac{f_B}{2f_\pi }%
\frac 1{1-q^2/M_B^2}  \label{33}
\end{eqnarray}
while $B^{*}$--pole contributions, using the parametrization (\ref{32}) $%
M_{B^{*}}\simeq M_B,$ $m_\pi ^2=0$, are 
\begin{eqnarray}
f_{+}^{B^{*}}\left( q^2\right) &=&\lambda \frac{f_B}{f_\pi }\frac 1{%
1-q^2/M_B^2}  \label{34} \\
f_{-}^{B^{*}}\left( q^2\right) &=&-\lambda \frac{f_B}{f_\pi }\left\{ 1+\frac 
1{1-q^2/M_B^2}\right\}  \label{35}
\end{eqnarray}
In Eq.(\ref{33}), 
\begin{eqnarray}
\lambda _c=\frac{4m_q/m_b}{\left( 1+\frac{m_q}{m_b}\right) \left[ 1-\frac{%
m_b^2}{M_B^2}\left( 1-\frac{m_q}{m_b}\right) ^2\right] }  \label{36}
\end{eqnarray}
We wish to point out that the continum contribution to the form factors
consisting of equal time commutator and valence quark contributions simulate
a $B-$meson pole $q^2-$dependence of form factors $f_{\pm }\left( q^2\right)
.$ This seems to follow a general result in quark annihilation model \cite
{r27} of decay of a pseudoscalar meson into, for example, two photons when
one photon is off mass shell. The $q^2$ dependence in this case is that of
the pseudoscalar meson pole involved. Here both the currents are conserved.
In our case the axial vector current to which $\pi -$meson is coupled is
partially conserved and this is reflected in the constant $f_B/f_\pi $ in
the valence quark contributions given in Eqs. (\ref{24}) and (\ref{25}). But
this is exactly cancelled by the equal time commutator contribution so that $%
q^2$ dependence of the continum contribution is given by $B-$meson pole,
which is indistinguishable from the usual vector $B^{*}-$meson pole
contribution to $f_{\pm }\left( q^2\right) $ in the heavy quark symmetry
limit $\left( M_{B^{*}}\simeq M_B\right) $.

The effect of continum contribution as defined above seems to change the
parameter $\lambda $ in $B^{*}$ contribution to an effective one $\lambda
_{eff}=\lambda +\lambda _c/2$ so that 
\begin{eqnarray}
f_{+}\left( q^2\right) &=&\frac{f_B}{2f_\pi }\left[ \lambda _c+2\lambda
\right] \frac 1{1-q^2/M_B^2}  \label{36a} \\
f_{-}\left( q^2\right) &=&-\frac{f_B}{2f_\pi }\left[ \lambda _c+2\lambda
\right] \left\{ 1+\frac 1{1-q^2/M_B^2}\right\}  \label{36b}
\end{eqnarray}
These formulae, as already noted, hold in the chiral limit i.e. for $q^2$ in
the range determined by $E_\pi =\left( m_B^2+m_\pi ^2-q^2\right) /\left(
2m_B\right) \leq 1$GeV or for $q^2\geq 17$ GeV$^2$.

\section{Chiral symmetry constrained model for $f_{+}\left( q^2\right) $ in
the entire momentum transfer}

In order to impliment the constraint given in Eq.(\ref{36a}), we use the
extrapolation function \cite{r27new} 
\begin{eqnarray}
f_{+}\left( q^2\right) =\frac{f_{+}(0)}{1-a\ q^2/M_B^2+b\left(
q^2/M_B^2\right) ^2}  \label{n38}
\end{eqnarray}
which involves three parameters $f_{+}(0)$, $a$ and $b$. For $E_\pi \simeq
m_\pi \rightarrow 0$, or $q^2\simeq M_B^2$, Eq.(\ref{n38}) should reduce to
Eq.(\ref{36a}) which gives 
\begin{eqnarray}
1-a+b &=&0  \label{n39} \\
f_{+}(0) &=&\frac{f_B}{2f_\pi }\left( \lambda _c+2\lambda \right) \left(
a-2b\right)  \nonumber \\
&=&\frac{f_B}{2f_\pi }\left( \lambda _c+2\lambda \right) \left( 1-b\right)
\label{n40}
\end{eqnarray}
Thus the pole at $q^2=M_B^2$ is factored out in Eq.(\ref{n38}) and we obtain 
\begin{eqnarray}
f_{+}\left( q^2\right) =\frac{f_{+}(0)}{\left( 1-q^2/M_B^2\right) \left(
1-b\ q^2/M_B^2\right) }  \label{n41}
\end{eqnarray}
It is interesting to note that Eq.(38) implies that in the heavy mass and
large $E_\pi $ ($\gg 1$ GeV or $q^2\ll 17$ GeV$^2$) limit, $f_{+}$ behaves
like $1/E^2$ in agreement with that found earlier in the HQET-LEET (large
energy effective theory) formalism for heavy-to-light form factors \cite
{r29new}. The Eq.(\ref{n41}) suggests that we may interpret the second
factor in Eq.(\ref{n41}) as arising from a second pole at $q^2=M_{B^{\prime
}}^2$, where $M_{B^{\prime }}$ is some effective mass, so that 
\begin{eqnarray}
b=\frac{M_B^2}{M_{B^{\prime }}^2}  \label{n42}
\end{eqnarray}
and 
\begin{eqnarray}
f_{+}\left( 0\right) =\frac{f_B}{2f_\pi }\left[ \lambda _c+2\lambda \right]
\left( 1-\frac{M_B^2}{M_{B^{\prime }}^2}\right)  \label{n43}
\end{eqnarray}
It is tempting to interpret that the suppression factor $\left(
1-M_B^2/M_{B^{\prime }}^2\right) $ in Eq.(\ref{n43}) and the second pole in
Eq.(\ref{n41}) as arising from radial excitations of $B^{*}$. Then making
use of the formula of Ref. \cite{r28}, we find $M_{B^{\prime }}/M_B\simeq
1.14$ so that $b\simeq 0.77$. To proceed further so as to obtain numerical
estimates we make the following choice of other parameters: $%
M_{B^{*}}\approx M_B\simeq 5.28$ GeV, $m_q/m_b=0.063$ and $m_b=4.757$ GeV 
\cite{r29}, which give $\lambda _c\simeq 0.826$.

We need also the values for $f_B$ and $\lambda $ which have considerable
uncertainity. We use $f_B=0.187$ GeV and $\lambda =0.5$ for our numerical
predictions. Then from Eq.(\ref{n40}), we obtain $f_{+}(0)=0.30$. With the
above choice of parameters, we plot the form factor of Eq.(\ref{n41}) in
Fig. 2. Also shown for comparison are the $B^{*}$ pole contribution given in
Eq.(\ref{34}), as well as the predictions obtained respectively from
light-cone sum rules \cite{r24} and on the light front \cite{r23}. In Fig. 3
we give the comparison of our prediction for $f_{+}\left( q^2\right) $ with $%
f_B=0.150$ GeV and those of \cite{r5} and \cite{r6} as well as that of a
recent calculation [32] where the $B-\pi $ transition form factors are
obtained for the whole range of qsquare by using a different method of
iterpolation between small and large values of $q^2$. This figure also gives
a comparison to lattice QCD data [33]. In Fig. 4, we plot the pion momentum
distribution in units of $\left| V_{ub}\right| ^2$ while Fig. 5 gives the
comparison of our prediction for this distribution with $f_B=0.150$ GeV and
those in refrences \cite{r5} and \cite{r6}. In each case we also give this
distribution for the $B^{*}$ pole for comparison which shows that $B^{*}$
pole is a good approximation to the full form factors upto $E_\pi $ of $\sim
1$ GeV.

Finally we calculate the branching ratio for $B\rightarrow \pi l\nu $ using
the form factors given in Eq.(\ref{n41}) and $f_{+}(0)$ in Eq.(\ref{n43})
with our choice of parameters, given previously. Our prediction is $\Gamma
=13.7\left| V_{ub}\right| ^2$ ps$^{-1}$ so that using $\tau _B=1.56$ ps we
obtain for the branching ratio 
\begin{eqnarray}
{\cal B}\left( B^0\rightarrow \pi ^{-}l^{+}\nu \right) \simeq 21.4\left|
V_{ub}\right| ^2  \label{n44}
\end{eqnarray}
To indicate sensitivity of this result on values of $\lambda $ and $f_B$
which we take $0.3\leq \lambda \leq 0.7$ and $0.150<f_B<0.187$ GeV, we can
express our prediction as 
\begin{eqnarray}
{\cal B}\left( B^0\rightarrow \pi ^{-}l^{+}\nu \right) \simeq \left( 20.0\pm
11.5\right) \left| V_{ub}\right| ^2  \label{n45}
\end{eqnarray}
With the CLEO measurement \cite{r30}, the result given in Eq.(\ref{n44})
means 
\begin{eqnarray}
\left| V_{ub}\right| =\left( 2.90\pm 0.48\right) \times 10^{-3}  \label{n46}
\end{eqnarray}
This is consistent with the result from exclusive decays quoted by the
Particle Data Group \cite{r31}: 
\[
\left| V_{ub}\right| =\left( 3.3\pm 1.1\right) \times 10^{-3} 
\]
and that obtained from the inclusive decays $\left| V_{ub}/V_{cb}\right|
=0.08\pm 0.02$ with $V_{cb}=0.0395$ : $\left| V_{ub}\right| =\left( 3.2\pm
0.8\right) \times 10^{-3}.$

\section{Conclusion}

We have presented a framework in which constraints from chiral symmetry and
quark model are used to predict the form factors for the exclusive $%
B\rightarrow \pi ^{+}l^{-}\nu $ decay in the entire physical range of
momentum transfer. The valence quark contribution is calculated in the
chiral symmetry approach. This togather with the equal time commutator
contribution simulate a B-meson pole $q^2$--dependence of the form factors.
This and $B^{*}$ pole contribution being obtained in chiral symmetry, are
valid for $E_\pi \leq 1$ GeV. This constraint is then implimented through
the extrapolation function for $f_{+}\left( q^2\right) $. The resulting form
factor, valid for the entire physical range for $q^2$, represents a
softening of $B^{*}$ pole behavior and the supression of the chiral coupling
which can be interpreted as being provided by the radial excitation of $B$.

The shape of the pion momentum distribution shown in Figs. 4 and 5 should be
able to distinguish our model from the others. The predicted branching ratio
for $B\rightarrow \pi l\nu $ in unit of $\left| V_{ub}\right| ^2$ is
sensitive to the values of $\lambda $ and $f_B$ as examplified in Eq.(\ref
{n45}), once we select the $b$ quark mass and the ratio $m_q/m_b.$ We
emphasize that uncertanities in the predicted branching ratio for $%
B\rightarrow \pi l\nu $ are due to uncertainities in the external parameters 
$\lambda $ and $f_B$ and are not intrinsic to the model.

With the CLEO measurement of the branching ratio for $B\rightarrow \pi l\nu $
, our prediction (\ref{n44}) with $\lambda =0.5$ and $f_B=0.187$ GeV give $%
\left| V_{ub}\right| $ as in Eq.(\ref{n46}) which is consistent with the
value quoted by the Particle Data Group. With some better knowledge of $%
\lambda $, $f_B$ and $m_b,$ the errors in form factors and the branching
ratio can potentially be reduced.

\section{Acknowledgements}

Two of us (R and Al-Aithan) wish to acknowledge the support of King Fahd
University of Petroleum and Minerals for this work while (Gilani) thanks the
Abdus Salam International Center for Theoretical Physics for support. We
also thank Prof. T. Feldmann for helpful comments and informing us of
references \cite{r29new} and \cite{r32new}.

\section{Figure Captions}

\begin{enumerate}
\item[Figure 1]  Valence--quark contribution.

\item[Figure 2]  The form factor $f_{+}\left( q^2\right) $ as a function of
the momentum transfer $t=q^2$. The solid curve is our prediction for $%
\lambda =0.5$ and $f_B=0.187$ GeV. The dashed line is the $B^{*}-$pole
contribution as given in Eq. (30). The dotted and dashed--dotted lines are
respectively the predictions of Refs. [23] and [24].

\item[Figure 3]  Same as those in Figure 2 but for $f_B=0.150$ GeV. The
dashed--dotted, dotted and dash-dot-dot lines are respectively the
predictions of Refs. [5], [6] and [32]. The data points correspond to
lattice results [33].

\item[Figure 4]  The pion energy distribution in units of $\left|
V_{ub}\right| ^2$as function of pion energy. The dashed line is $B^{*}$
prediction.

\item[Figure 5]  Same as in Fig. 4 for $f_B=0.150$ GeV. The dashed--dotted
and dotted lines are respectively the predictions of Ref. [5] and [6].
\end{enumerate}

\end{document}